\documentclass[final, twocolumn, a4paper]{aastex63}

\usepackage{color}
\usepackage{enumitem}
\usepackage{hyperref}
\usepackage{verbatim}
\usepackage{amsmath,amstext,mathrsfs}
\usepackage[all]{hypcap} 
\usepackage{afterpage}    
\usepackage{marginnote}
\usepackage{makecell}
\usepackage{subfigure}
\usepackage{multirow}
\usepackage{lineno}

\shorttitle{Curved magnetic field in NGC 628 }
    \shortauthors{Zhao et al}
\usepackage{enumitem}
\setlist[enumerate]{listparindent=\parindent}
\makeatletter

\newcommand{\Rmnum}[1]{\expandafter\slowromancap\romannumeral #1@}
\makeatother

\begin{document}
    \hfuzz = 150pt
\title{\Large\bfseries Magnetic Field of Molecular Gas Measured with the Velocity Gradient Technique  II: Curved Magnetic Field in kpc-Scale Bubble of NGC\,628 }

\correspondingauthor{Mengke Zhao, Jianjun Zhou}
\email{mkzhao628@gmail.com, zhoujj@xao.ac.cn}

\author[0000-0003-0596-6608]{Mengke Zhao}
\affil{Xinjiang Astronomical Observatory, Chinese Academy of Sciences,  Urumqi, 830011, People's Republic of China}
\affil{University of Chinese Academy of Sciences, Beijing, 100049, People's Republic of China}

\author[0000-0003-0356-818X]{Jianjun Zhou}
\affil{Xinjiang Astronomical Observatory, Chinese Academy of Sciences,  Urumqi, 830011, People's Republic of China}
\affil{Key Laboratory of Radio Astronomy, Chinese Academy of Sciences  Urumqi,830011, People's Republic of China}
\affil{Xinjiang Key Laboratory of Radio Astrophysics, Urumqi 830011, People's Republic of China}

\author[0000-0003-3389-6838]{Willem A. Baan}
\affil{Xinjiang Astronomical Observatory, Chinese Academy of Sciences,  Urumqi, 830011, People's Republic of China}
\affil{Netherlands Institute for Radio Astronomy ASTRON, 79901 PD Dwingeloo, the Netherlands}

\author[0000-0002-8455-0805]{Yue Hu}
\affiliation{Department of Physics, University of Wisconsin-Madison, Madison, WI 53706, USA}
\affiliation{Department of Astronomy, University of Wisconsin-Madison, Madison, WI 53706, USA}
\author[0000-0002-7336-6674]{A. Lazarian}
\affiliation{Department of Astronomy, University of Wisconsin-Madison, Madison, WI 53706, USA}
\affiliation{Centro de Investigación en Astronomía, Universidad Bernardo O’Higgins, Santiago, General Gana 1760, 8370993, Chile}

\author[0000-0002-4154-4309]{Xindi Tang}
\affil{Xinjiang Astronomical Observatory, Chinese Academy of Sciences,  Urumqi, 830011, People's Republic of China}
\affil{Key Laboratory of Radio Astronomy, Chinese Academy of Sciences  Urumqi,830011, People's Republic of China}
\affil{Xinjiang Key Laboratory of Radio Astrophysics, Urumqi 830011, People's Republic of China}

\author{Jarken Esimbek}
\affil{Xinjiang Astronomical Observatory, Chinese Academy of Sciences,  Urumqi, 830011, People's Republic of China}
\affil{Key Laboratory of Radio Astronomy, Chinese Academy of Sciences  Urumqi,830011, People's Republic of China}
\affil{Xinjiang Key Laboratory of Radio Astrophysics, Urumqi 830011, People's Republic of China}
\author[0000-0002-8760-8988]{Yuxin He}
\affil{Xinjiang Astronomical Observatory, Chinese Academy of Sciences,  Urumqi, 830011, People's Republic of China}
\affil{Key Laboratory of Radio Astronomy, Chinese Academy of Sciences  Urumqi,830011, People's Republic of China}
\affil{Xinjiang Key Laboratory of Radio Astrophysics, Urumqi 830011, People's Republic of China}
\author{Dalei Li}
\affil{Xinjiang Astronomical Observatory, Chinese Academy of Sciences,  Urumqi, 830011, People's Republic of China}
\affil{Key Laboratory of Radio Astronomy, Chinese Academy of Sciences  Urumqi,830011, People's Republic of China}
\affil{Xinjiang Key Laboratory of Radio Astrophysics, Urumqi 830011, People's Republic of China}
\author{Weiguang Ji}
\affil{Xinjiang Astronomical Observatory, Chinese Academy of Sciences,  Urumqi, 830011, People's Republic of China}]
\author{Zhengxue Chang}
\affil{College of Mathematics and Physics, Handan University, No.530 Xueyuan Road, Hanshang District, 056005 Handan, PR China}
\author{Kadirya Tursun}
\affil{Xinjiang Astronomical Observatory, Chinese Academy of Sciences,  Urumqi, 830011, People's Republic of China}
\affil{Key Laboratory of Radio Astronomy, Chinese Academy of Sciences  Urumqi,830011, People's Republic of China}
\affil{Xinjiang Key Laboratory of Radio Astrophysics, Urumqi 830011, People's Republic of China}

\begin{abstract}

We report the detection of the ordered alignment between the magnetic field and kpc-scale bubbles in the nearby spiral galaxy, NGC\,628. 
Applying the Velocity Gradient Technique (VGT) on CO spectroscopic data from the ALMA-PHANGS, the magnetic field of NGC\,628 is measured at the scale of 191\,pc ($\sim$ 4\,$''$).
The large-scale magnetic field is oriented parallel to the spiral arms and curves around the galactic bubble structures in the mid-infrared emission observed by the James Webb Space Telescope (JWST). 
Twenty-one bubble structures have been identified at the edges of spiral arms with scales over 300\,pc, which includes two kpc-scale structures. 
These bubbles are caused by supernova remnants and prolonged star formation and are similar to the outflow chimneys found in neutral hydrogen in galactic disks.
At the edge of the bubbles, the shocks traced by the OIII emission present a curved magnetic field that parallels the bubble's shell.
The magnetic field follows the bubble expansion and binds the gas in the shell to trigger further star formation.
By analyzing the larger sample of 1694 bubbles, we found a distinct radial-size distribution of bubbles in NGC\,628 indicating the star formation history in the galaxy.

\end{abstract}


\section{introduction}

Magnetic fields are a primary component of the interstellar media (ISM) that play an important role in star formation, and the evolution of spiral galaxies \citep{2007ARA&A..45..565M,2013pss5.book..641B}.
Magnetic fields have been observed over the past decades at all spatial scales in the Galaxy and in extra-galactic sources using far-infrared and radio observations \citep{1994RPPh...57..325K, 2004NewAR..48.1289B,2015A&ARv..24....4B,2020ApJ...888...66L,2020AJ....160..167J}.
Shocks, supernovae, star formation activities, and other physical processes may affect and distort the local magnetic field structures in the ISM (eg: \citealt{2004MNRAS.353..550B,2017NatAs...1E.158L,2021A&A...647A..78A} and the structure of spiral arms and spurs \citep{1980ApJ...242..528E,1985ApJ...297...61B,2012ApJ...756...45L,2014ApJ...792..122L}.

The recent introduction of the Velocity Gradient Technique (VGT;\,\citealt{2017ApJ...835...41G,2018ApJ...853...96L,2018MNRAS.480.1333H}) provides a new way to measure the magnetic field using Doppler-shifted emission lines 
in nearby galaxies 
\citep{2023ApJ...946....8T,2022ApJ...941...92H,2023MNRAS.519.1068L}. 
In a turbulent magnetized medium, the turbulent eddies will extend along the local B-field and have a velocity gradient perpendicular to their semi-major axis \citep{1995ApJ...438..763G,1999ApJ...517..700L}.
These velocity gradients of elongated eddies are expected to be perpendicular to the local B-field orientation.
The PHANGS-ALMA survey provides high-resolution CO isotopolog data where anisotropies of the turbulence can be extracted and used to trace the velocity gradients of eddies. 
The ability of the VGT technique to trace molecular-gas-associated magnetic fields has been tested in observations in nearby galaxies \citep{2022ApJ...941...92H,2023MNRAS.519.1068L}.

Early Release Observations (ERO; \citealt{2022ApJ...936L..14P}) of the spiral galaxy NGC 628 with the James Webb Space Telescope (JWST) reveal new features compared with previous observations in the form of 'bubble structures' along spiral arms that have no mid-infrared continuum emission.
Such features resemble the chimney outflows found in neutral hydrogen related to star formation in edge-on spiral galaxies. 
Combined with the observation of the magnetic field, these bubbles provide new understanding of how the magnetic field may affect the structure of the spiral galaxy.

In this work, we aim to study the magnetic field in and around the bubbles structures of the spiral galaxy NGC\,628 at the distance of 7.3\,Mpc \citep{2004AJ....127.2031K}, which exhibits ongoing star formation but is not undergoing a central starburst \citep{2018PASP..130h4101L}. 
The details of the archival data are shown in Section\,\ref{sect2}.
In Section\,\ref{sect3}, we introduce the details of the VGT and VDA (Velocity Decomposition Algorithm; \citealt{2021ApJ...910..161Y}) techniques.
The magnetic field structure and galactic bubbles in NGC\,628 are described and discussed in Sections\,\ref{sect3} and \ref{sect4}.
We discuss the nature of the bubbles and their magnetic field structure in Section\,\ref{sect5}. 
A summary has been provided in Section\,\ref{sect6}.

\begin{figure*}
    \centering
    \includegraphics[width=16cm]{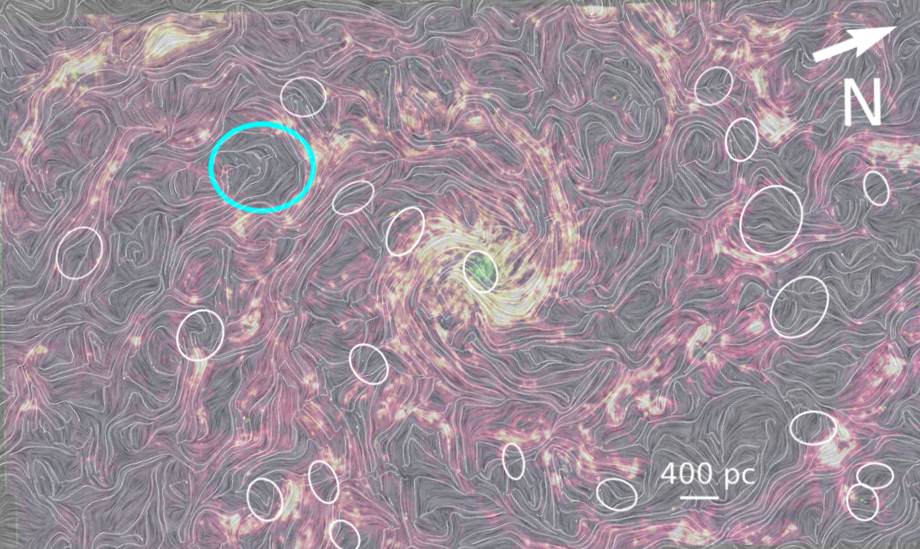}
    \caption{The curved magnetic field aligns with the bubble structures in NGC\,628. 
    This map displays the magnetic field structure measured with VGT for the CO emission of the nearby spiral galaxy NGC\,628, where the magnetic field orientation is indicated as a line-integral-convolution (LIC) map.
    The background is the RGB image at the mid-infrared with red: F1130W; green: F1000W; blue: F770W.
    The cyan circle is the one-kpc-scale bubble.
    The white circles show the position of 20 bubbles with scale sizes $>$ 390\,pc.
    }
    \label{fig1}
\end{figure*}

\section{The NGC\,628 Data}\label{sect2}

\subsection{The spectral line emission}

In this work, we select $^{12}$CO\,(2-1) emission as the main tracer to measure the magnetic field of the spiral galaxy NGC628 with the VGT technique. 
The spectral cube data comes from the PHANGS-ALMA survey \citep{2021ApJS..257...43L}, which includes 90 nearby galaxies (d $\leq$  20\,Mpc). 

high signal-to-noise ratio data with the RMS noise level of $\sim$ 0.30\,$\pm$\,0.13 K\,km\,s$^{-1}$.
The spatial resolution for NGC\,628 is $\sim$ 1.12$''$ (FWHM $\sim$ 53.5 pc) and the velocity resolution is around 2.5 km\,s$^{-1}$ \citep{2021ApJS..257...43L}. 
The rms level of the PHANGS-ALMA Survey is around 0.3 $\pm$ 0.17 K\,km\,s$^{-1}$.

In addition, the OIII emission in NGC\,628 has been selected to trace the shocks, which comes from the PHANGS-MUSE Survey \citep{2022A&A...659A.191E} with a sub-arcsec beam scale size.

\subsection{The continuum emission}

The continuum emissions of NGC\,628 are observed by the James Webb Space Telescope (JWST) Early Release Observations (ERO; \cite{2022ApJ...936L..14P}).
The filters F770W (7.7\,$\mu$m), F1000W (10\,$\mu$m), F1130W (11.3\,$\mu$m), and F2100W (21\,$\mu$m) from the JWST Mid Infrared Instrument (MIRI) will be used in this work. 

\subsection{The polarization data} 

The synchrotron polarization of NGC\,628 has been observed with the $\it{Karl G. Jansky}$ Very Large Array (JVLA) at S -band (2.6–3.6 GHz effective bandwidth and spatial resolution $\sim$ 18$''$; \citealt{2017A&A...600A...6M}).
The orientation of the magnetic field $\theta_B$ derived from the JVLA polarization data are estimated by: 
\begin{equation}
    \begin{aligned}
        \rm \theta_B = \frac{1}{2}\,arctan(U,Q) \, + \frac{1}{2}\pi,
    \end{aligned}
\end{equation}
where the Q and U are the Stokes parameters. 

\section{Using the VGT Method}\label{sect3}
\subsection{The anisotropy of the MHD turbulence}

The velocity gradient technique (VGT; \citealt{2017ApJ...835...41G,2018ApJ...853...96L,2018MNRAS.480.1333H}) is the main analysis tool used in this work. 
This has been developed based on the current theories of MHD turbulence \citep{1995ApJ...438..763G} and fast turbulent re-connection \citep{1999ApJ...517..700L}.  These studies revealed that turbulent eddies are anisotropic, which means that the eddies are elongated along the local magnetic field lines \citep{1995ApJ...438..763G,1999ApJ...517..700L}. 
If the scale of the eddies is decomposed into parallel ($l_\parallel$) and perpendicular  ($l_\bot$) components with respect to the magnetic field, the anisotropy suggests $l_\bot \ll l_\parallel$.
This property has been confirmed with numerical simulations \citep{2000ApJ...539..273C,2001ApJ...554.1175M,HXL21} and with solar wind observations \citep{2016ApJ...816...15W,2020FrASS...7...83M,2021ApJ...915L...8D}.
Another important property of MHD turbulence is that the velocity fluctuations are scale-dependent such that \citep{1999ApJ...517..700L}:
\begin{equation}
	\label{eq.v}
	v_l\simeq v_{\rm inj}(\frac{l_\bot}{L_{\rm inj}})^{\frac{1}{3}}M_{\rm A}^{1/3},
\end{equation}
where $v_l$ is the fluctuation at scale $l$, $v_{\rm inj}$ is the velocity at the injection scale $L_{\rm inj}$, and $M_A$ is Alfv\'en Mach number. Together with the anisotropic relation $l_\bot \ll l_\parallel$, the scaling of the velocity gradients can be easily obtained \citep{2020ApJ...898...65Y}:
\begin{equation}
	\label{eq.grad}
	\begin{aligned}
		\nabla v_l&\propto\frac{v_{l}}{l_\bot}\simeq \frac{v_{\rm inj}}{L_{\rm inj}}(\frac{l_{\perp}}{L_{\rm inj}})^{-\frac{2}{3}} M_{\rm A}^{\frac{1}{3}}.
	\end{aligned}
\end{equation}
Eq.~\ref{eq.grad}  suggests that the gradient amplitude increases when the scale $l_\bot$ decreases, i.e., the small resolved eddies correspond to strong velocity gradients. Therefore, in extragalactic sources, where the 
shear velocity and the differential rotation might contribute to the total velocity gradient, the contribution from the MHD turbulence would dominate at small scales ($<100$~pc), as demonstrated in \citet{2022ApJ...941...92H,2023MNRAS.519.1068L}.

\subsection{The Velocity Gradient Technique}\label{A.VGT}

 The input of the VGT technique is the high-resolution position-position-velocity (PPV) data cube of the $^{12}$CO emission in this work. 
The extraction of the velocity channel Ch(x,y) information from the PPV cube of the spectral line data has been done via the velocity caustics effect. 
The concept of the velocity caustics effect has been defined as the effect of the density structure distorting the turbulence and shear velocities along the line of sight \citep{2000ApJ...537..720L}. 
The density structure changes significantly for the different velocity channel and the resulting velocity fluctuations are thought to be most prominent in thin channel maps \citealt{2000ApJ...537..720L,2021ApJ...915...67H}. 
The difference for thin and thick channels is as follows: 
\begin{equation}\label{eq5}
    \begin{aligned}
        \Delta v \, > \, \sqrt{\delta (v^2)}, \rm thick\,channel,
    \end{aligned}
\end{equation}\label{eq6}
\begin{equation}
    \begin{aligned}
        \Delta v \, < \, \sqrt{\delta (v^2)}, \rm thin\,channel,        
    \end{aligned}
\end{equation}
where $\Delta v$ is the velocity channel width and $\sqrt{\delta (v^2)}$ is the velocity dispersion. 

The thin velocity channel map Ch$_i$(x,y) is applied on the VGT technique \citep{2022MNRAS.511..829H} to extract the velocity information from the PPV cubes via the velocity caustics effect.
Each thin velocity channel map Ch$_i$(x,y) is used to calculate the pixelised gradient map $\psi^i_g (x,y)$: 
\begin{equation}\label{eq.a4}
    \begin{aligned}
        \rm\bigtriangledown_x Ch_i(x,y) = Ch_i(x,y) - Ch_i(x-1,y) \,,
    \end{aligned}
\end{equation}
\begin{equation}\label{eq.a5}
    \begin{aligned}
        \rm\bigtriangledown_y Ch_i(x,y) = Ch_i(x,y) -Ch_i(x,y-1)\,, 
    \end{aligned}
\end{equation}
\begin{equation}
    \begin{aligned}
        \rm\psi^i_g (x,y) = tan^{-1} (\frac{\bigtriangledown_y Ch_i(x,y)}{\bigtriangledown_x Ch_i(x,y)} \,,
    \end{aligned}
\end{equation}
where $\bigtriangledown_x$Ch$_i$(x,y) and $\bigtriangledown_y$Ch$_i$(x,y) are the x and y components of the gradient, respectively. 
The (x,y) two-dimensional position will be displayed in the (RA,DEC) plane of the sky for pixels whose spectral line emission has a signal-to-noise ratio greater than 3, which is high enough for credibility.

The orientation of the magnetic field is perpendicular to the velocity gradient direction in each sub-region.
A sub-block averaging \citep{2017ApJ...837L..24Y} method has been used to export the velocity gradients from the raw gradients within a sub-block of interest and then plotted in a corresponding histogram of the raw velocity gradient orientations $\psi^i_g$.
The size of the sub-block is set as 20\,$\times$\,20 pixels, which determines the size of the final magnetic field resolution.
Using sub-block averaging, the total n$_v$ processed gradient maps $\psi^i_{g_s}(x,y)$ with i = 1,2,...,n$_v$ are taken to calculate the pseudo-Stokes-parameters $Q_g$ and $U_g$.
Then, these pseudo-Stokes-parameters $Q_g$ and $U_g$ of the gradient-inferred magnetic field would be constructed by:
\begin{equation}
    \begin{aligned}
        \rm Q_g(x,y) = \sum\limits_{i=1}^{n_v} Ch_i(x,y)cos(2\psi^i_{g_s}(x,y)),
    \end{aligned}
\end{equation}
\begin{equation}
    \begin{aligned}
        \rm U_g(x,y) = \sum\limits_{i=1}^{n_v} Ch_i(x,y)sin(2\psi^i_{g_s}(x,y)),
    \end{aligned}
\end{equation}
\begin{equation}\label{eq12}
    \begin{aligned}
        \rm\psi_g = \frac{1}{2}tan^{-1}\frac{U_g}{Q_g},
    \end{aligned}
\end{equation}
where I$_i$(x,y) is the two dimensions integrated intensity of spectra cubes and n$_v$ is the number of velocity channels. 
The $\psi_g$ pseudo polarization is derived from the pseudo-Stokes-parameters $Q_g$ and $U_g$, which projects the three-dimensional $\psi^i_{g_s}$ data into a two-dimensional pseudo polarization angle $\psi_g$.
The pseudo magnetic field orientation $\psi_B$ is perpendicular to the pseudo polarization angle $\psi_g$ on POS: 
\begin{equation}\label{eq13}
    \psi_B = \psi_g + \pi /2 ,
\end{equation}
This pseudo magnetic field orientation is the B-field orientation measured with the VGT method from the spectral line data.
Applying the VGT technique on the $^{12}$CO emission from the PHANGS-ALMA survey \citep{2021ApJS..257...43L}, the magnetic field orientations can be measured and be presented as an LIC \citep{Cabral_lic} as shown in Figure\,\ref{fig1}.
The resolution of the B-field measured with VGT can be approaching 4$"$ ($\sim$ 190\,pc).

\subsection{The Velocity Decomposition Algorithm}\label{B.VDA}

The VGT technique is based on the position-position-velocity statistics \citep{2000ApJ...537..720L}, where the PPV cube has two components: a velocity and a density contribution. 
While the VGT technique is more sensitive to the velocity contribution,
the Velocity Decomposition Algorithm (VDA) is a new technique that separates the velocity and density contribution from the PPV cube \citep{2021ApJ...910..161Y},.
By this method, the accuracy of VGT tracing of the magnetic field will be improved \citep{2022ApJ...934...45Z}.

In this work, the gas traced by the CO emission may come from star formation regions tracing the spiral arms.
These star formation regions may have supersonic motions and the properties of the velocity component $V(X, v, \Delta v)$ may be calculated as:
\begin{equation}
    \begin{aligned}
        \rm V(X, v, \Delta v) = - c^2_s \frac{\partial Ch(X, v, \Delta v)}{\partial v} \,,
    \end{aligned}
\end{equation}
where Ch(X, v, $\Delta$v) is the channel of the PPV cube, X means the position, v is the local velocity, and $\Delta$v is the velocity channel width. 
The velocity of sound $c_s$ can be calculated by assuming a uniform temperature $\sim$ 10 K, which results in a value $c_s$ $\sim$ 186 m s$^{-1}$.
Using these velocity properties $p_v$ to replace the values for Ch(x,y) in Eq.\,\ref{eq.a4} and \ref{eq.a5}, this VGT-VDA method can trace the magnetic field better. 
The VDA algorithm exhibits a sensitivity to spectral lines characterized by a high signal-to-noise ratio, particularly those present in structures of high density and small scale. 

\begin{table}[]
    \centering
    \caption{The position and the supernova remnant numbers of the bubbles in NGC\,628}
    \begin{tabular}{c c c c c }
    \hline
    \hline
        ID & RA & DEC & size & Estimated \\
           &    &     & (pc) & SNR number \\
    \hline
    1 & 1h36m44.64s & 15d46m20.7 s & 1164  & $\sim$ 194 \\
    2 & 1h36m40.84s & 15d48m09.6s & 808  & $\sim$ 135  \\
    3 & 1h36m35.95s & 15d48m06.1s & 418  & $\sim$ 70 \\
    4 & 1h36m36.29s & 15d48m11.1s & 408  & $\sim$ 68 \\
    5 & 1h36m37.39s & 15d48m01.4s & 520  & $\sim$ 87 \\
    6 & 1h36m37.51s & 15d47m12s & 460  & $\sim$ 77 \\
    7 & 1h36m38.47s & 15d46m07.9s & 392  & $\sim$ 65 \\
    8 & 1h36m38.63s & 15d46m51.85s & 408  & $\sim$ 68 \\
    9 & 1h36m39.33s & 15d48m08.5s & 762  & $\sim$ 127 \\
    10 & 1h36m39.43s & 15d46m07.8s & 496 & $\sim$ 83 \\
    11 & 1h36m39.5s & 15d45m53.2s & 490  & $\sim$ 82 \\
    12 & 1h36m40.71s & 15d48m35.7s & 394  & $\sim$ 66 \\
    13 & 1h36m40.98s & 15d46m27.8s & 500  & $\sim$ 83 \\
    14 & 1h36m42.25s & 15d48m09.8s & 484  & $\sim$ 81 \\
    15 & 1h36m42.41s & 15d45m52.7s & 592  & $\sim$ 99 \\
    16 & 1h36m42.81s & 15d46m47.1s & 588  & $\sim$ 98 \\
    17 & 1h36m43.25s & 15d48m08.1s & 476  & $\sim$ 79 \\
    18 & 1h36m43.64s & 15d46m38.4s & 512  & $\sim$ 85 \\
    19 & 1h36m44.39s & 15d45m32.7s & 614  & $\sim$ 102 \\
    20 & 1h36m45.49s & 15d46m35.8s & 510  & $\sim$ 85 \\
    \hline
    \hline
    \end{tabular}
    \tablecomments{These are 20 large-scale bubbles. The scale of Bubble 1 is at one kpc and others are hundreds-pc scale.
    The ID of large-scale bubbles is shown in the 1st column.
    The position is shown in the 2nd and 3rd columns.
    The position and size of bubbles are defined from the JWST MIRI continuum data.
    The SN number shows the estimated number of SNRs required to make a bubble of this size (see Section \ref{sect5.2}).
    }
    \label{tab1}
\end{table}

\subsection{VGT method in Shock Fronts}

On smaller scales, the shocks induce velocity jumps that are perpendicular to shock fronts, which can interfere with the turbulence-induced gradients.
While the effect of shocks would reduce with a decreasing velocity channel width, it should be noticeable for the CO data having a velocity channel width large enough at $\sim$ 2.5 km s$^{-1}$.
At a shock front, the velocity perpendicular to the shock will be reduced while the velocity component parallel to the shock front will be unchanged. 
For oblique shocks, this changes the orientation of the velocity vector and the velocity gradient at the shock front \citep{2019ApJ...886...17H}. 
Since the anisotropy of the MHD turbulence does not affect the shock, a magnetic field component in the same direction as the propagation direction of the shock front will still suggest a velocity gradient perpendicular to the magnetic field, and a propagation direction of the shock perpendicular to the shock front. 
For a magnetic field that is perpendicular to the propagation direction of the shock front or has some angle, the relative orientation of the velocity gradient and magnetic field is not maintained at 90 degrees \citep{2019ApJ...878..157X}. 
For a shock front that is parallel to the magnetic field, the direction of the magnetic field and the velocity gradient tend to be parallel.
In cases where the shock front is oriented perpendicularly to the local magnetic field, the magnetic field that is unaffected by the local shock front aligns perpendicularly to the orientation of the mean velocity gradient (see Eq.\,\ref{eq13}). 
This physical process mirrors that observed in the VGT pipeline, as detailed in Section \ref{A.VGT}.
Conversely, when the shock front aligns paralleling the local magnetic field, the magnetic field influenced by the shock aligns paralleling  the mean velocity gradient orientation and perpendicularly to the pseudo-magnetic field orientation
\begin{equation}\label{eqfrontparal}
    \begin{aligned}
        & if \, B \, \mathop{//}\, {\rm shock front},\,\, \psi_{Bs} = \psi_B + 90^\circ, \\
    \end{aligned}
\end{equation}
\begin{equation}
    \begin{aligned}
        if \, B \, \bot \, {\rm shock front}, \,\, \psi_{Bs} = \psi_B,\\
    \end{aligned}
\end{equation}
where the $\psi_B$ is the VGT orientation and $\psi_{Bs}$ is the magnetic field orientation at the shock front.
The shock fronts can thus be traced by the velocity field of the OIII emission. 
The local magnetic field orientation can be discerned from the polarization at lower resolution, which encompasses the shock front.

\section{Result}\label{sect4}

\subsection{The magnetic fields in NGC\,628}

The magnetic fields in the galaxy NGC\,628 have been measured using the Velocity Gradient Technique using the high-resolution $^{12}$CO (2-1) spectral line emission. 
Using the VGT technique, the magnetic field is derived at a scale of 191\,pc ($\sim$ 4$''$) with the sub-block size set at 20$\times$20 pixels and with a pixel size of the $^{12}$CO emission of 0.2$''$.
The magnetic field structure of NGC\,628 is shown in Figure\,\ref{fig1}. 

On larger scales, the magnetic field is distributed along the spiral arms and the B-field orientations are nearly parallel to the tangential direction of the spiral arms. 
On smaller scales, the magnetic field displays curved structures around the spiral arms (ring-like shapes).

\begin{figure*}
    \centering
    \includegraphics[height= 7cm]{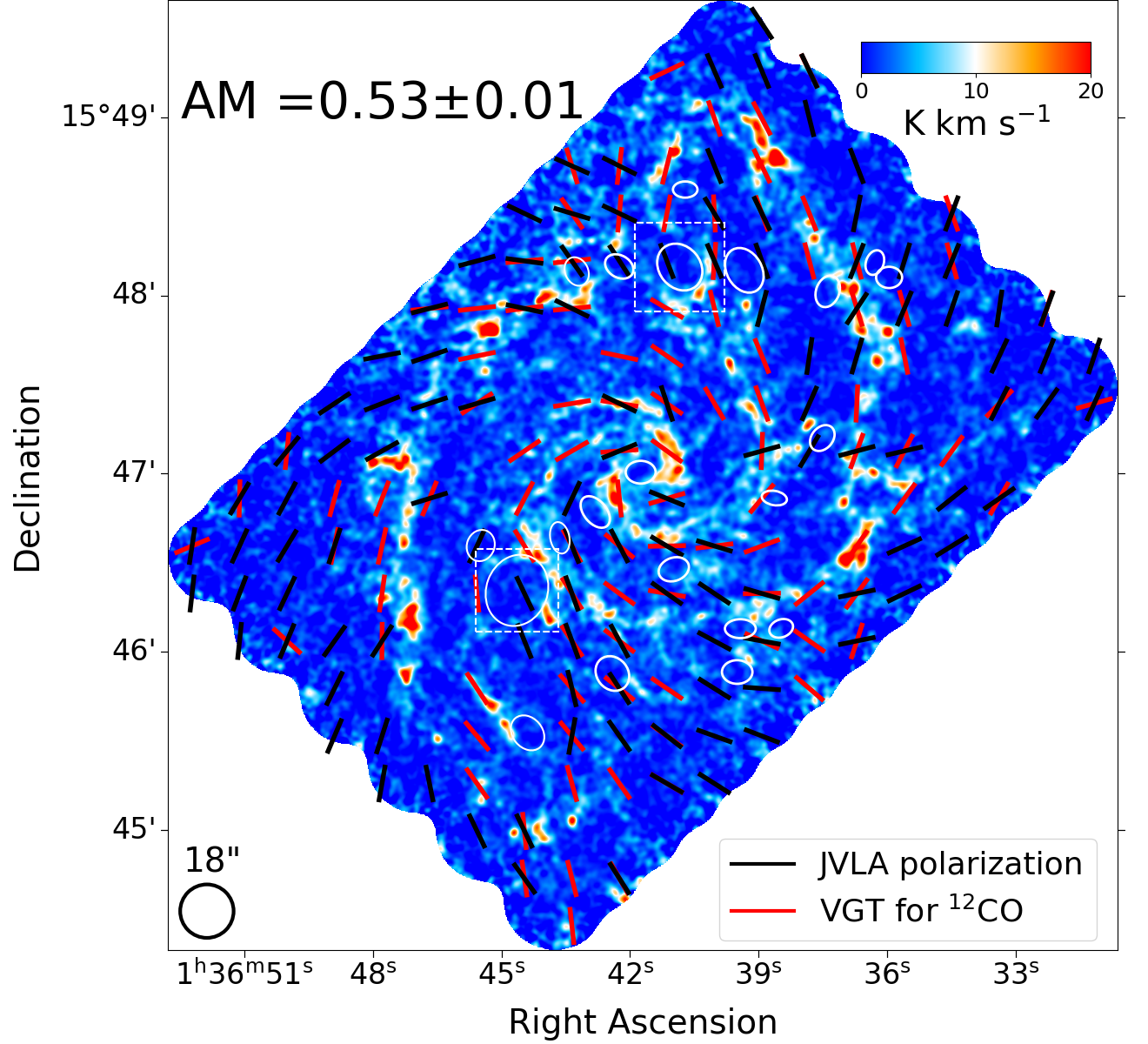}
    \includegraphics[height= 7.4cm]{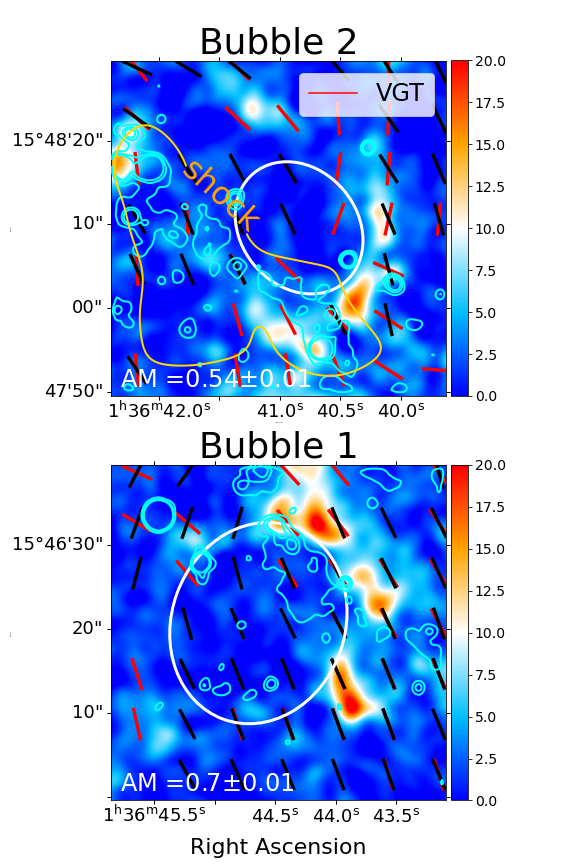}
    \includegraphics[height= 7.4cm]{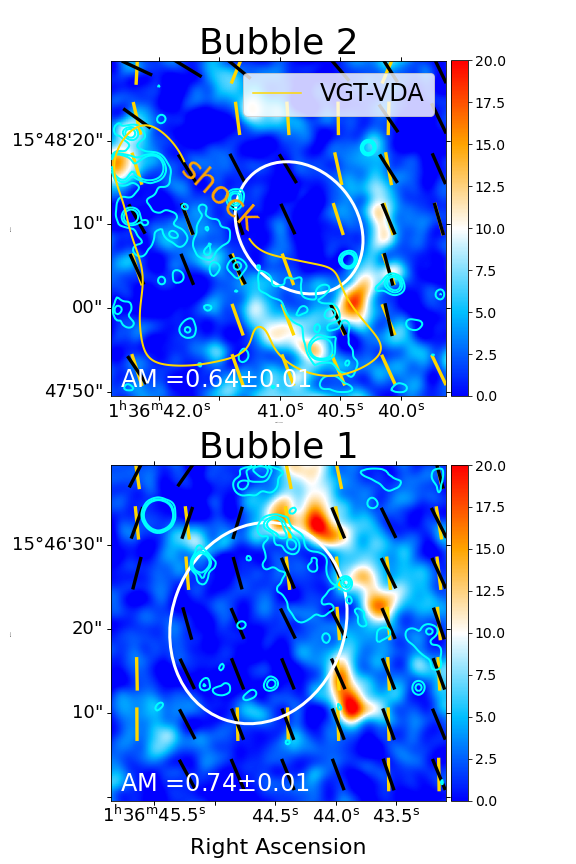}
    \caption{The magnetic field distribution measured with VGT and taken from the JVLA synchrotron polarization. 
    The red and black vectors in the left panel display the magnetic field orientation measured with VGT and the JVLA polarization at 3.1 GHz \citep{2017A&A...600A...6M}, respectively. 
    The background is the $^{12}$CO (2-1) emission intensity map. 
    The mean value for the Alignment Measure AM is 0.53 with an uncertainty of 0.01. 
    The middle panel shows the distribution of the magnetic field at the two kpc-scale bubbles (see Fig.\,\ref{fig1}) associated with the white boxes in the left panel. 
    The cyan contours show the flux of OIII emissions at [4,\,5,\,6]$\times$10$^{-16}$ cm$^{-2}$\,erg\,s$^{-1}$ as determined from the PHANGS-MUSE survey data \citep{2022A&A...659A.191E}. 
    The OIII lines overlap the large region of Bubble\,2 and are located in a small part of Bubble\,1
    The orange contour shows the shock front identified by smoothing the OIII emissions to 18$''$.
    The right panels display the comparison of magnetic field orientations between the VGT-VDA and JVLA polarization at 3.1 GHz.
    The yellow vectors show the magnetic field orientation measured by VGT-VDA.}
    
    \label{fig2}
\end{figure*}

\subsection{Bubbles in extragalactic sources}

The JWST Early Release Observations \citep{2022ApJ...936L..14P} present details of the spiral galaxy NGC\,628.
The high-resolution mid-infrared emission from JWST, characterized by a Full Width at Half Maximum (FHWM) of approximately 10 pc, presents a unique opportunity for investigating bubbles within extragalactic environments. 
This advanced capability allows for the observation of distinctive bubble features such as shells, stellar sources, and the hot emissions encapsulated by these bubbles \citep{2006ApJ...649..759C,2014ApJS..212....1A,2019MNRAS.488.1141J}.
In a recent study, \citealt{2023ApJ...944L..24W} identified numerous bubbles within the extragalactic system NGC 628, demonstrating a broad range of scales with diameters spanning from 12 to 1164 pc. 
From the bubble catalog of \citealt{2023ApJ...944L..24W},we have identified 21 sizable bubbles exceeding twice the beam width of the magnetic field measured with VGT (approximately 390 pc). 
Of these, one particular bubble situated at the galactic center does not align with the focus of our work; it may be influenced more by the galactic center's black hole rather than stellar feedback and star formation activities. 
Our attention centers on the remaining 20 large-scale bubbles within our sample, which are potentially linked to local star formation activities, as detailed in Table \ref{tab1}.
These large-scale bubbles display the hole-like shape around the ring shape. 
The ring structure contains large amounts of gas and shocks traced by CO and O\,III emission (see Fig.\,\ref{fig6}).
The scale of our structures is all larger than 390\,pc, where 390\,pc is twice the beam for measuring the magnetic field structure accurately.
Details of the bubble structures are shown in Table\,\ref{tab1}.

\subsection{Comparison of the B-fields from VGT and from Synchrotron Polarization}\label{C.JVLA}

The magnetic field measured with VGT for CO emission has a higher resolution ($\sim$ 4$''$), compared with previously determined B-field structures derived from synchrotron polarization ($\sim$ 18$''$) with JVLA at 3.1 GHz \citep{2017A&A...600A...6M}.
We smooth the magnetic field derived by VGT to the same scale as that from JVLA resolution ($\sim$ 18$''$) and compare the two versions of the magnetic field. 
The alignment between the B-field orientations measured with polarization $\theta_B$ and VGT $\Phi_B$ is quantified as the \textbf{Alignment Measure} (AM; \citealt{2017ApJ...835...41G}): 
\begin{equation}\label{AMeq}
    \begin{aligned}
        \rm AM = 2(\left \langle cos^2(\theta_B - \Phi_B) \right \rangle - \frac{1}{2}) \,.
    \end{aligned}
\end{equation}
The range of AM values should be from -1 to 1, where AM values close to 1 mean that $\phi_B$ is parallel to $\psi_g$ and AM values close to -1 indicate that $\phi_B$ is perpendicular to $\psi_g$. 
The uncertainty in the AM value $\sigma_{AM}$ may be given by a standard deviation divided by the square root of the sample size. 

\subsubsection{The B-field in the galactic disk}

Figure\,\ref{fig2} shows the magnetic field structure measured with VGT for CO emissions and from synchrotron polarization with a beam size of 18\,$''$.
The synchrotron radiation could originate from the warm ISM in star formation regions  \citep{2017A&A...600A...6M}, while the CO emission originates from the cold ISM in the disk \citep{2022ApJ...941...92H,2023MNRAS.519.1068L}. 

Star-forming regions may be actively generating cosmic rays (CRs) and the cooling times of these CRs emitting at 3.1\,GHz could be estimated as $t_{synch}$ $\sim$ 1.5($B$/mG)$^{-1.5}$(h$v$/kev)$^{-0.5}$ $\sim$ 13\,Myrs, where the total magnetic field strength is around 10$\mu$G \citep{2017A&A...600A...6M}.
The average diffusion time of $\sim$ 1 GeV CR protons in the Milky Way is around 10\,Myrs for an assumed diffusion coefficient of 10$^{28}$\,cm$^{2}$\,s$^{-1}$ and a halo size around 1\,kpc \citep{1976RvMP...48..161G,1990acr..book.....B,1998ApJ...509..212S,2019PhRvD..99j3023E}.
However, in star-formation regions and molecular clouds the CR diffusion is suppressed as a result of the CR steaming along confusing turbulence magnetic fields \citep{2022ApJ...927...94X} or mirror diffusion \citep{2021ApJ...923...53L,2021ApJ...922..264X}, which is also observed in Milky Way . 
Assuming that the disk height is hundreds of pc, the diffusion time could thus be one or two orders of magnitude longer.
Since the cooling time of CR electrons is less than the diffusion time, electrons at 3.1\,GHz would not diffuse into the halo. 
Therefore, CR electrons are still actively generated but are confined in the vicinity of the galactic disc and the origin of the CO emission could be similar to that of synchrotron emission.

Figure\,\ref{fig2} shows that the mean AM value is approaching 0.53 and is $\geq$ 0.5, which suggests that the magnetic field orientations from VGT are nearly parallel to those inferred from synchrotron polarization.
The magnetic field derived by VGT is basically consistent with the previous results obtained from synchrotron polarization.
Because the magnetic field measured with VGT for CO emission has a higher resolution in the galactic disk, the results may reflect the magnetic field at the spiral arms but also the small-scale structures around it.

\begin{figure*}
    \centering
    \includegraphics[width = 14cm]{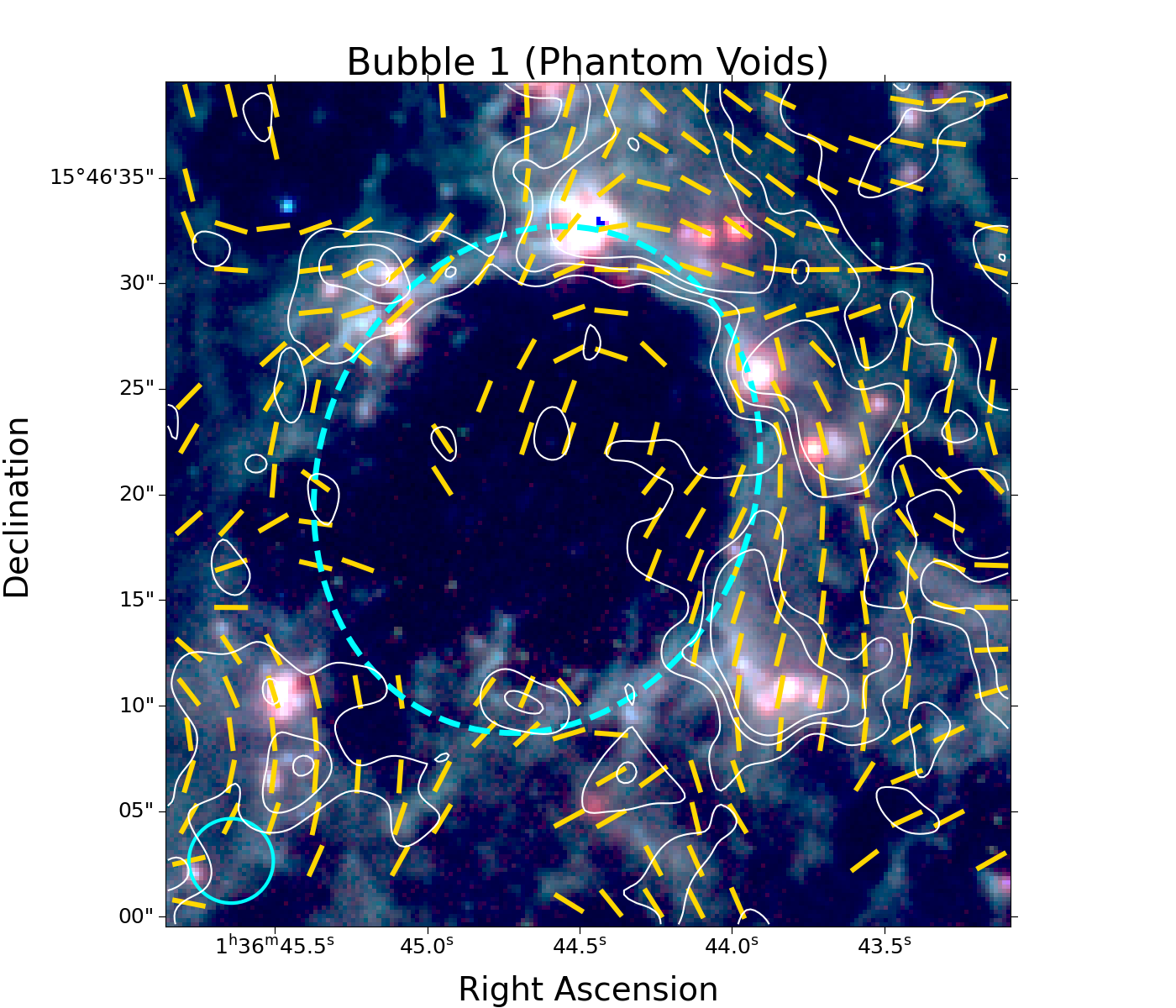}
    \caption{The magnetic field structure measured with VGT-VDA for kpc-scale Bubble in NGC\,628. 
    The panels display the curved magnetic field at bubble\,1.
    The beam size is about 4$''$ and is displayed as a cyan circle in the left bottom corner.
    The image background is a three color map in the mid-infrared with red: F770W, green: F1130W, and red: F2100W.
    The yellow vectors represent the magnetic field orientations measured by the VGT-VDA technique for $^{12}$CO. 
    The white contours are the contour levels of the $^{12}$CO integrated intensity at 5, 10, and 15$\sigma$.
    The cyan line shows the bubble structure.}
    \label{fig3}
\end{figure*}

\begin{figure*}
    \centering
    \includegraphics[width = 14cm]{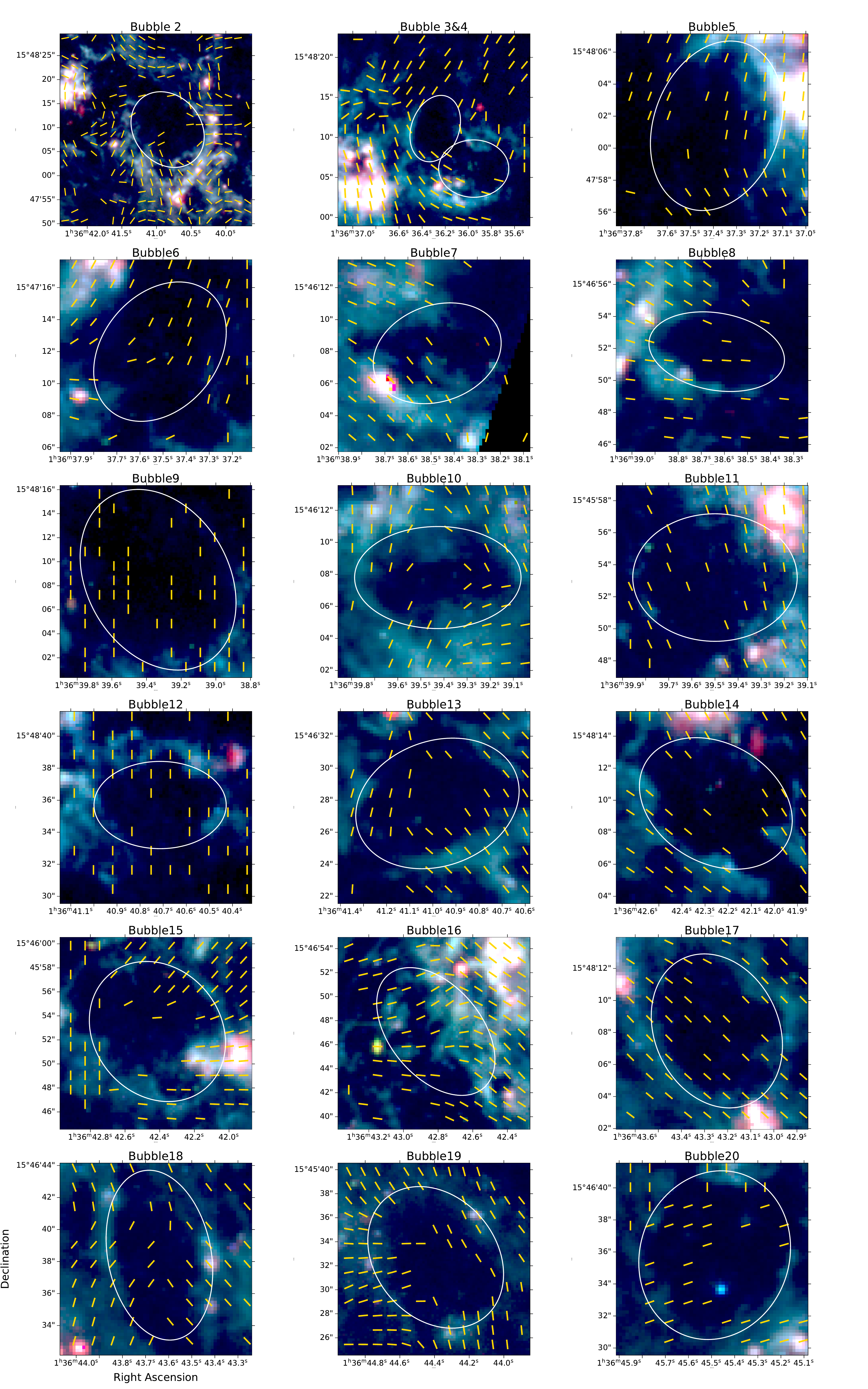}
    \caption{Magnetic field structure of large-scale bubbles in NGC\,628 measured with the VGT-VDA method using the CO emission. 
    The yellow vectors display the magnetic field orientations.
    The background is an RGB image with red: F2100W, green: F1130W, and blue: F770W. 
    The left bottom corner displays the size of beam ($\sim$ 4$''$). }
    \label{figsample}
\end{figure*}

\subsubsection{The B-field at the Bubbles}

The two largest bubbles, Bubble 1 and Bubble 2 (elaborated in Table \ref{tab1}), exceed the spatial resolution of previous synchrotron polarization observations (FWHM $\sim$ 800\,pc) \citep{2017A&A...600A...6M}. 
These two large-scale bubbles offer exemplary instances for assessing the congruence between magnetic field orientation and VGT within minute-scale structures.
The OIII emission from PHANGS-MUSE Survey is the tracer of shocks and Figure\,\ref{fig2} shows that the shock in Bubble\,1 is close but does not overlap the gas shell of the bubble. 
Because of the shock front and the gas not being in the same position, the shock could not change the velocity gradient of the cold neutral medium traced by the CO emission. 
By comparison between the magnetic field for VGT and that for polarization, the mean AM is 0.70 ($\pm$ 0.01). 
This means that the magnetic field measured with VGT could be similar to that from polarization.
In the no-shock region, the velocity gradient remains perpendicular to the magnetic field.

Different from Bubble\,1, a shock exists at the gas shell of Bubble\,2. 
As Figure\,\ref{fig2} shows, part of the bubble shell in Bubble\,2 coincides with a shock front and the region has massive OIII emission.
The OIII emission may be smoothed to trace the shock front at a scale of 18$''$, where the intensity is over 3$\times$10$^{-16}$ cm$^{-2}$\,erg\,s$^{-1}$.
Here the shape of the shock front is parallel to the magnetic field derived by polarization.
The VGT orientation in this region is re-rotated by 90 degrees to achieve the B-field orientation, as dictated by Eq.\,\ref{eqfrontparal}. 
By re-rotating the VGT orientation at the shock front, the AM between two measures of magnetic fields is 0.54, and the two types of magnetic fields are parallel to one another.
In the shock front, the accuracy of VGT tracing magnetic field has been tested and the velocity gradient is close to being parallel to the magnetic field.
In the region of the shock front, the VGT orientation needs to be re-rotated 90$^\circ$.

For the two largest bubbles, we also use the VDA method to improve the accuracy of the VGT technique by removing the density contribution in the PPV data cube. 
As Fig.\,\ref{fig3} shows, we compare the magnetic field measured with VGT-VDA and 3.1\,GHz polarization. 
The AM value of Bubble\,1 and Bubble\,2 is up to 0.74 and 0.64, which is larger than the AM of VGT (Bubble\,1 = 0.7 and Bubble\,2 = 0.54).
This means that the VGT-VDA has higher accuracy in tracing magnetic fields than only the VGT method, but it also verifies that the magnetic fields measured with VGT for the CO emission. 
It also means that we can reliably use the VGT technique to study the small-scale structures in NGC\,628.

\section{Discussion}\label{sect5}

\subsection{Large-scale Bubbles as Supernova Remnants}

The bubble structures appear rather common around the inner spiral arms of NGC\,628 as seen in  Fig.\,\ref{fig1}.  
To establish the nature of the structure of these large-scale bubbles and the role of the magnetic field, the VGT-VDA method has been used to study the effects of the high-resolution magnetic field on large-scale bubble structures.
This method improves the accuracy of the VGT method as the application of VDA removes the effect of the density contribution to the velocity gradient, which is particularly important at the edges of the structures.

Supernova remnants (SNRs, \citealt{1988ARA&A..26..145T, 1988ARA&A..26..295W}) and prolonged star formation are the most likely cause for the creation of a cavity within the circular bubble structures.  
Supernova explosions could have enough energy to cause large-scale bubbles with scales up to hundreds of pc or even kpc.
These bubbles are found to be surrounded by gas structures with much-triggered star formation activities as traced by 11.3 $\mu$m and 21 $\mu$m emissions. 
This makes the bubbles to be relics of past star formation that only now show up at mid-infrared frequencies (see Fig.\,\ref{fig1}).

\begin{figure}
    \centering
    \includegraphics[width = 8.5cm]{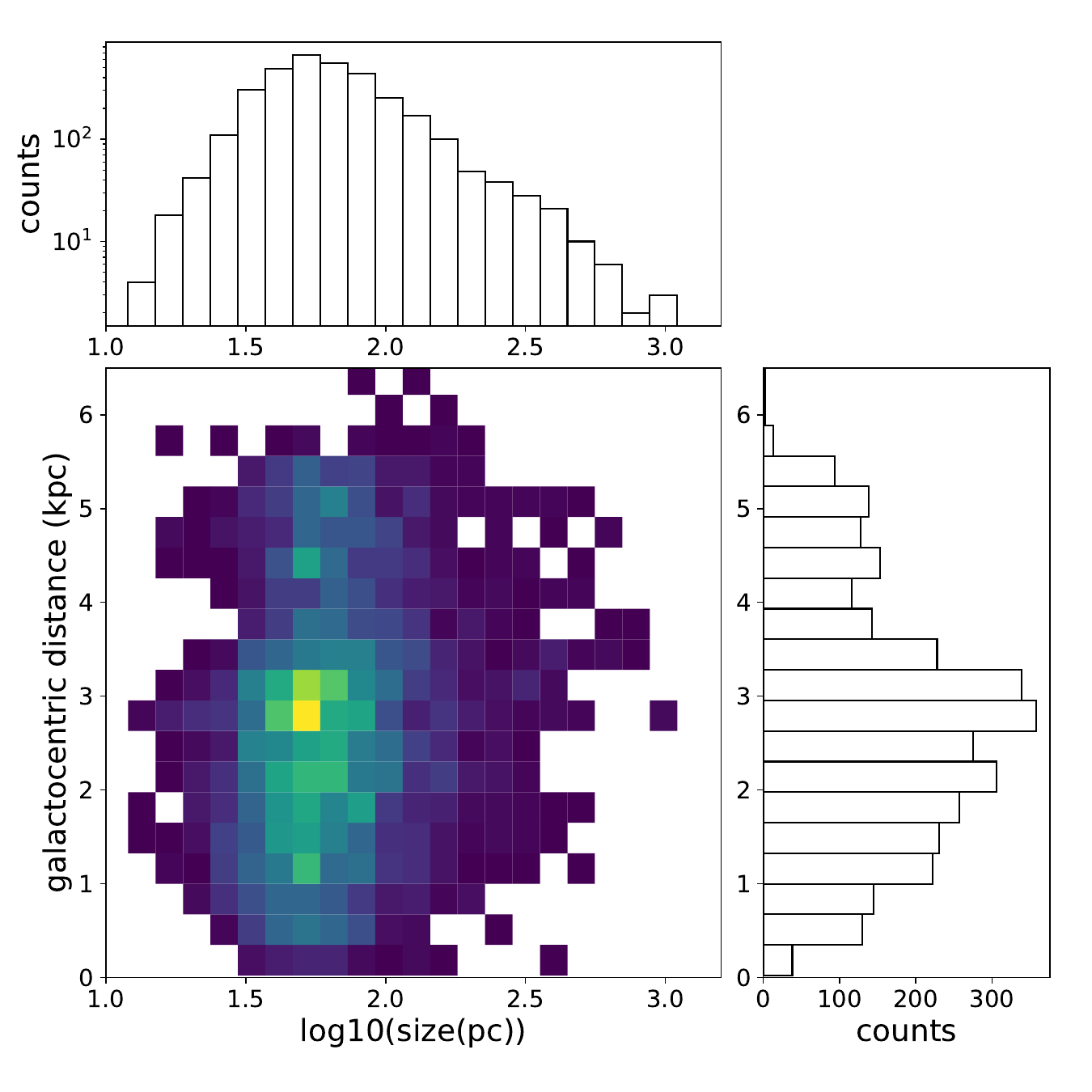}
    \caption{The distribution of the size of the bubbles versus galactocentric distance to the galaxy center using a probability density function and a 2D histogram. }
    \label{fig4}
\end{figure}

The kpc-scale Bubble\,1 in Figure\,\ref{fig3} may be used as a high-confidence example of a bubble, where the shape and magnetic field structure could be caused by SNRs.
The OIII emission serves as a valuable indicator for tracing the impact of shocks, as shown in Fig.\,\ref{fig2}.
At the edge of Bubble\,1, the shock front traced by the OIII emission is close to the bubble's shell but it does not overlap with the spiral arms traced by the CO emission.
The magnetic field morphology in the curved structure of Bubble\,1 is like a closed ring that is aligned around the shape of the bubble.
Due to the shock front being close to the shell at the edge of the bubbles, the supernovae create the shock and compress the gas to cause the bubbles and the shell. 
As a result of the gas flow towards the shell, the magnetic field has been distorted and forced to be perpendicular to the direction of matter flow (see Fig.\,\ref{fig4}). 
The curved magnetic field structure therefore represents the shocked boundaries of the large-scale bubble and serves as indirect evidence that the bubble structure and its curved magnetic field would be caused by the SNRs. 

Alternative star formation processes, such as stellar feedback and the expansion of HII regions, may exhibit a greater inclination toward instigating the smaller-scale structure of ISM. 
These formations might lack the requisite energy to give rise to structures on a scale of hundreds of pc.
Similar phenomena may also manifest within sub-regions of the large-scale bubbles. 
As an illustration, the orientations of magnetic fields in the sub-region of Bubble\,2 (see Fig. \ref{figsample}) are observed to be perpendicular to the gas structures in sub-regions with existing star formation as traced by 21$\mu$m emissions. 
It is noteworthy that local magnetic fields may undergo distortion due to the influence of star formation activities in this specific region, a phenomenon akin to those observed within the Milky Way \citep{2017NatAs...1E.158L,2024ApJ...961..124Z}.

\subsection{Bubble Size and Galactic Distribution}\label{sect5.2}

In this paper, only the large-scale bubbles have been considered and their magnetic field structures have been determined (see Fig.\,\ref{fig4}).
A more complete sample of bubbles in NGC\,628 includes 1694 bubbles with different sizes and is provided by \cite{2023ApJ...944L..24W}.
A histogram of the size distribution and the galacto-centric distance of the bubbles in this extended sample is presented in Fig.\,\ref{fig4} and shows a peak in the extended distribution at a bubble size of 50\,pc and at a galacto-centric distance of 3\,kpc. 
In the galactic distribution of bubbles, the number of bubbles increases nearly linearly up to the galacto-centric distance as 3\,kpc, and then they decrease sharply and non-linearly. 
The large-scale bubbles considered are thus mainly distributed in the inner part of the galaxy within 3\,kpc.

The bubble number distribution along the galactocentric distance is similar to the star formation rate distribution in the Milky Way \citep{2022ApJ...941..162E}.  The bubble scale of $\sim$ 50\,pc could be typical for NGC\,628 and is similar to the largest supernova remnants found in the Milky Way \citep{1988RvMP...60....1O,2021A&A...648A..30B}. 
However, this typical bubble size in NGC\,628 is much larger than the typical size of $\sim$ 6\,pc for most SNRs in the Milky Way and it is unknown how many SNs actually contributed to this large size.
The peak radial position of the bubble distribution exists at about 1/6 to 1/5 of the disk size and suggests a ring structure with an enhanced star formation history. 
The existence of this ring requires further investigation because of its implications for the radial density structure and the rotation curve in the disk, as well as for the location of the Inner Lindblad resonance in the galaxy.

\begin{figure*}
    \centering
    \includegraphics[width = 14cm]{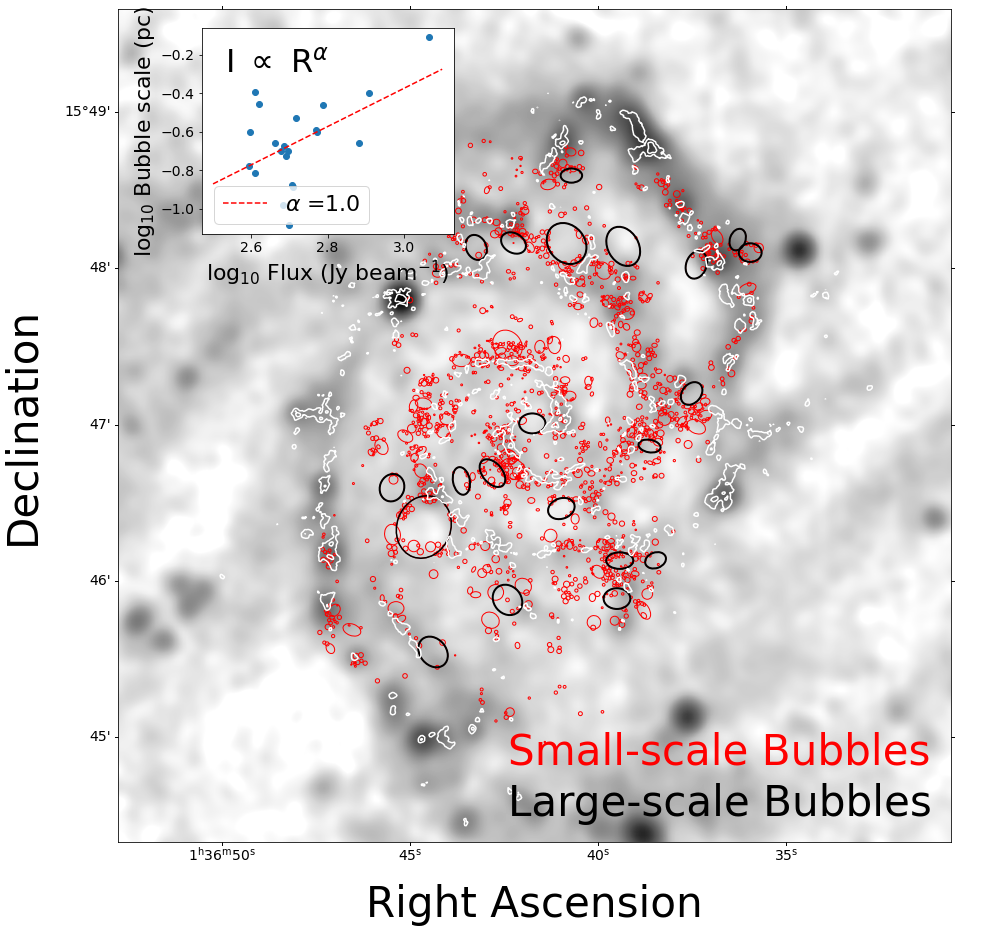}
    \caption{\textbf{Distribution of bubbles aligned with synchrotron radiation at 3.1 Ghz.} 
    The main panel shows the distribution of bubbles in NGC\,628 \citep{2023ApJ...944L..24W}.
    The black rings display the large-scale bubbles (FWHM $\textgreater$ 390 pc) and the red rings show the small-scale bubble(FWHM $\textless$ 390 pc) in this work.
    The background is mapping the 3.1 GHz emissions at the scale of 8$''$ \citep{2017A&A...600A...6M}.
    The white dash line displays the intensity map of $^{12}$CO (2-1) emissions.
    The sub-panel shows the distribution between scale R and flux I of large-scale bubbles.
    The red line is the fitting Eq.\,\ref{eqIR}.}
    \label{figIR}
\end{figure*}

\subsection{The origin of large-scale bubbles}

The large-scale bubbles could be caused by the superposition of multiple supernova explosions for three reasons:  \\
1. The shock front traced by the OIII emissions is distributed around the shell of the bubbles (see Fig.\,\ref{fig2}). \\
2. There is triggered star formation at the edges of the bubbles, which is traced by the 11.3$\mu$m and 21$\mu$m emission (see Fig.\,\ref{fig1} and \ref{fig3}). \\
3. The curved magnetic field structure is aligned with the structure of the bubbles.

The size of the large-scale bubbles ($>$ 300\,pc) considered here clearly exceeds the size of even the largest SNR in the Milky Way ($\sim$\,50 pc).
Such large-scale bubble structures may only result from massive star formation and the contribution of multiple supernova explosions. 
The shock wave from an SN could compress the surrounding gas and merge with shells from many others \citep{2023ApJ...944L..24W} to form a large-scale bubble with a surface close to spherical.
Depending on the thickness of the disk, the bubble cavity will become cylindrical and form a chimney outflow perpendicular to the disk as seen in neutral hydrogen surveys. 
The diffuse gas in the disk would be compressed in a circular shell that is also constrained by the embedded local magnetic field and this enriched shell may lead to triggered star formation. 
A situation similar to that of NGC\,628 is found, for instance, in the galaxy NGC\,6946 \citep{2012ApJ...754L..35H}.

The radio emission from SNRs results from high-energy electrons accelerated during a supernova explosion and end up spiraling along the magnetic field structure \citep{2019JApA...40...36G,2024ApJS..270...21C}.
\citealt{2017A&A...600A...6M} provides a high-resolution synchrotron radiation map of NGC\,628 at 3.1\,GHz with resolution up to 8$''$ (FWHM $\approx$ 390\,pc; twice of B-field beam in this work).
As Figure\,\ref{figIR} shows, the synchrotron emission is aligned with the spatial distribution of bubbles from \citealt{2017A&A...600A...6M},which includes both the large-scale bubbles (FWHM $\textgreater$ 390 pc, the resolution of synchrotron radiation) and the small-scale bubbles (FWHM $\textless$ 390 pc).
These bubbles could be candidates of SuperNova Remnants.
The regions with richer synchrotron emissions have massive small-scale bubbles compared with the large-scale bubbles.
The small-scale bubbles are located at the spiral arms with rich gas.
In the spiral arms, despite impactful physical phenomena like supernova explosions, it proves challenging to propel massive amounts of gas and facilitate the formation of large-scale structures.
On the other hand, the diffuse gas surrounding the large-scale bubbles could be pushed by supernovae over greater lengths and form the large-scale structure of bubbles.

The diffuse gas and limited synchrotron emissions surrounding the expansive bubbles on a large scale imply that the number of supernovae contributing to the formation of such bubbles is comparatively modest when juxtaposed with equivalent areas in spiral arm regions.
Assuming that the aggregate flux, $I$, of the large-scale bubbles and their scale size, $R$, follow a power-law relation:
\begin{equation}\label{eqIR}
    I \propto R^\alpha \propto (\frac{R}{R_0})^{\alpha},
\end{equation}
where $R_0$ is the characteristic scale of one SNR, and $\alpha$ signifies the slope between $R$ and $I$ in log-log space.
When the flux within the bubble is uniform, the value of $\alpha$ becomes 2 (since the area is proportional to the square of the radius). 
By fitting the total flux and scale of each large-scale bubble, the derived alpha converges around 1 (less than 2). 
The growth rate of the flux in the bubbles is below that of the growth in scale. 
As the bubble's scale expands, the synchrotron emissions from SNR may undergo dilution.

We adopt a maximum size of a single SNR in the Milky Way ($\sim$ 6\,pc) as the characteristic scale populating the bubbles.
The SNR number of large-scale bubbles can be estimated by the scale $R$ of the bubbles:
\begin{equation}
    {\rm N (SNR)} = k(R / R_0) \propto (I/I_0)
\end{equation}
Because of the unknown characteristic flux $I_0$, we assume the factor k as 1.
Then the largest bubble, Bubble\,1 (scale $\sim$ 1164\,pc) 
contains an estimated 194 SNRs. 
Details regarding the number of SNRs in other large-scale bubbles can be found in Tab. \,\ref{tab1}. 
The observed large-scale bubbles in NGC\,628 provide a good picture of the succession of star formation and its history.

\section{Summary}\label{sect6}

The magnetic fields of NGC\,628 has been measured with the VGT technique using the $^{12}$CO (2-1) emission at the scale of 4$''$ (FWHM $\sim$ 191\,pc).
The magnetic field is in alignment with the structure of the bubbles in NGC\,628 and forms a surrounding ring-like structure leaving a central cavity at the Mid-infrared wavelengths. 

1. The high-resolution magnetic field measured with VGT for the CO emission is ordered and distributed along the spiral arms at the large scale. 
The result is in agreement with the preview observations of the synchrotron polarization at the low resolution of 18$''$. 
For smaller-scale bubble sources, the VGT-VDA method has been used to trace the magnetic field and improve accuracy, and the magnetic fields display curved structures. 

2. 20 large-scale bubbles (390\,pc -1164\,pc) in NGC 628 display a hole-like shape surrounded by a ring structure, which includes rich gas reservoirs and shock wave trails.
These large-scale bubbles are all distributed on the edge of spiral arms and are the result of multiple supernova activities.

3. The distribution of bubble size versus galactocentric distance of a large sample of 1694 bubbles displays a peak for a size around 50\,pc at a galactocentric distance of 3\.kpc. 
The number of bubbles in the galactic distribution increases linearly up to 3\,kpc and rapidly decreases beyond this distance.
The distribution of bubbles in NGC\,628 forms a ring structure that is densely filled with bubbles and reveals the intense star formation history in the central region of NGC\,628. 
The distribution of bubbles is similar to the star formation rate distribution in Milky Way.
The bubbles observed in NGC\,628 have been observed to be associated with outflow chimneys in neutral hydrogen in other galaxies.

4. The apparent ring of 50 pc bubble structures at about 3 kpc that surrounds regions of intense star formation in the inner part of the disk in NGC\,628 has
implications for the radial density structure the rotation curve in the disk, and the location of the Inner Lindblad resonance in the galaxy. 

5. The bubble structure appears to be caused by repeated supernova explosions and prolonged star formation, which squeezes gas towards the edge of the bubbles and triggers new star formation in these compressed regions. 
The organization of the magnetic field in the shock boundary surrounding the bubble structures will be perpendicular to the propagation direction of the SNR shocks and parallel to the shell of OIII gas.
While these bubbles are relics of past star formation, they continue to trigger new star formation.

6. The large-scale bubbles could be caused by the contribution of massive/multiple supernova explosions.  

The multiple supernovae drive the diffuse gas over a large length and cause the large-scale bubbles.
Scaling of the power needed to form such large-scale bubbles suggests contributions from large numbers (up to hundreds) of single SNR as observed in the Milky Way.
The number and size of the bubbles observed in NGC\,628 reveal the long-term star formation history in the galaxy, which is a common evolutionary process in galaxies.

Further plan are to observe the magnetic field using dust polarization at sub-arcsec resolution in order to study the small-scale magnetic field structure of the large-scale bubbles. 
Analysis of a face-on galaxy will promote understanding of similar physical processes in the Milky Way.

\section*{Acknowledgements}

We thank the referee for the careful reading of the paper and the constructive comments.
This work was mainly funded by the National Natural Science foundation of China (NSFC) under grant No.\,11973076. 
Partial funding was obtained from the CAS "Light of West China" Program under grant No.\,2020-XBQNXZ-017,
the NSFC under grant Nos.\,11903070, 12173075, 12103082, and 12303027, 
the Natural Science Foundation of Xinjiang Uygur Autonomous Region under grant No.\,2022D01E06, the Heaven Lake Hundred Talent Program of Xinjiang Uygur Autonomous Region of China, 
the Tianshan Innovation Team Plan of Xinjiang Uygur Autonomous Region (2022D14020), 
Xinjiang Key Laboratory of Radio Astrophysics (No. 2022D04033),
the Youth Innovation Promotion Association CAS, 
and the Project of Xinjiang Uygur Autonomous Region of China for Flexibly Fetching in Upscale Talents and the Hebei NSF No. A2022109001. 
A.L. and Y.H. acknowledge the support of the NASA ATP 80NSSC20K0542, AAH7546, and NASA TCAN 144AAG1967. 
W.A.B has been funded by the Chinese Academy of Sciences President’s International Fellowship Initiative grant No.\,2021VMA0008 and 2022VMA0019. 

\bibliographystyle{aasjournal}
\bibliography{reference}

\appendix





\section{A CO and OIII intensity map}

\begin{figure}[h]
    \centering
    \includegraphics[width=19cm]{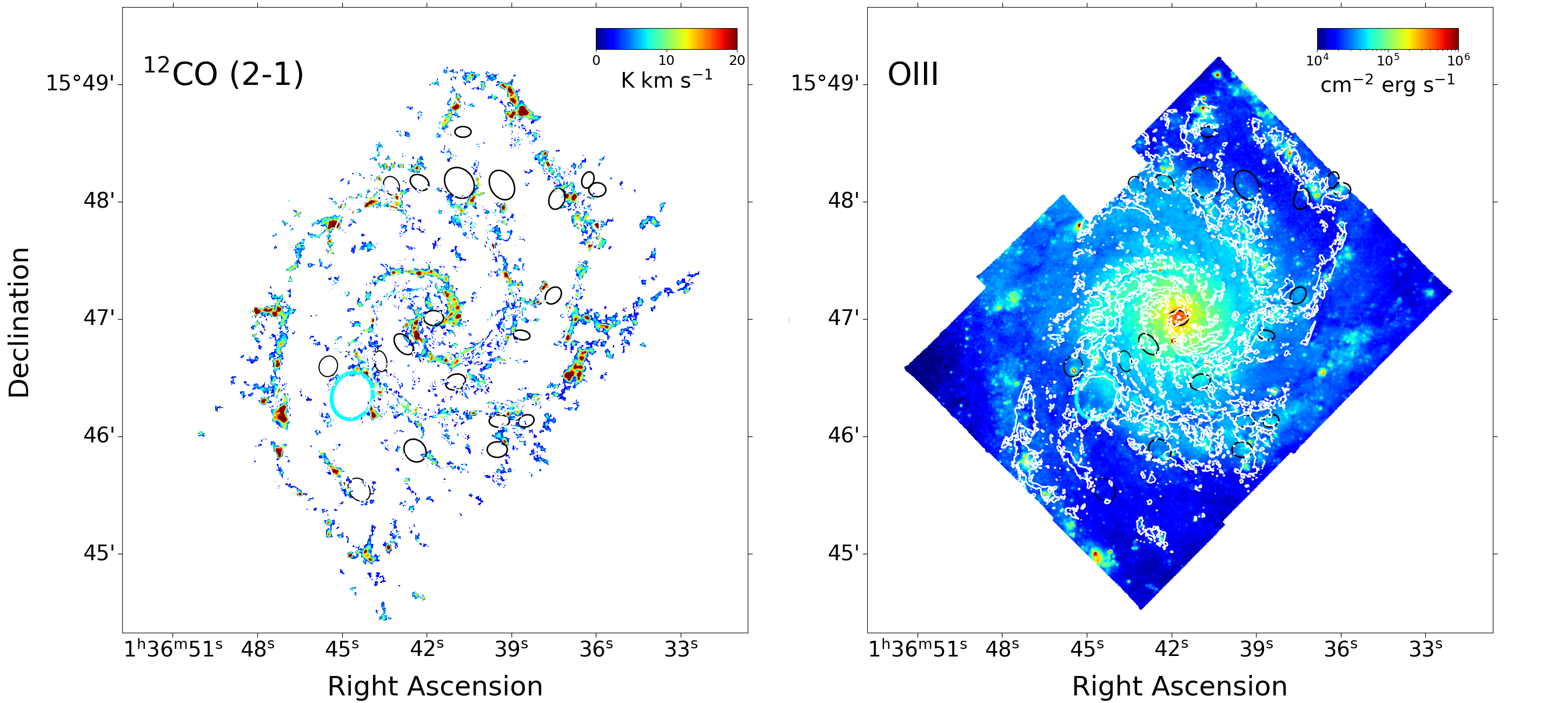}
    \caption{The intensity map of the $^{12}$CO\,(2-1) and OIII emissions, from left to right, respectively. 
    The white contours display the structure of the 11.3$\mu$m continuum observed by JWST MIDI (F1130W) at 47 MJy\,sr$^{-1}$. 
    The cyan and black circles display the position of the bubbles.}
    \label{fig6}
\end{figure}

Figure\,\ref{fig6} shows the intensity maps of the main tracers in this work $^{12}$CO\,(2-1) and OIII.

\end{document}